\documentclass[12pt,english,floatfix,superscriptaddress,aps,prd,preprint,showkeys]{revtex4}
\usepackage[utf8]{inputenc}
\usepackage{amsmath}
\usepackage{amssymb}
\usepackage{amsbsy}
\usepackage{amsfonts}
\usepackage{amsopn}
\usepackage{amstext}
\usepackage{graphicx}
\usepackage{amssymb}
\usepackage{amsfonts}
\usepackage{amsmath}
\usepackage{amsmath,amsthm,amsfonts,amssymb}
\usepackage[mathcal]{eucal}
\usepackage{mathrsfs}
\usepackage{graphicx}
\usepackage[english]{babel}
\usepackage{color}
\usepackage{esint}
\usepackage[dvips]{epsfig}
\usepackage[dvips]{graphicx}
\usepackage{float}
\usepackage{units}
\usepackage{textcomp}

\usepackage{hyperref}
\usepackage{slashed}

\newcommand{\ie}{\begin{equation}}
\newcommand{\fe}{\end{equation}}
\newcommand{\se}{\begin{eqnarray}}
\newcommand{\ff}{\end{eqnarray}}

\begin{document}

\title{Thermodynamic properties of the noncommutative Dirac oscillator with a permanent electric dipole moment}

\author{R. R. S. Oliveira}
\email{rubensrso@fisica.ufc.br}
\affiliation{Universidade Federal do Cear\'a (UFC), Departamento de F\'isica,\\ Campus do Pici, Fortaleza - CE, C.P. 6030, 60455-760 - Brazil.}


\author{R. R. Landim}
\email{renan@fisica.ufc.br}
\affiliation{Universidade Federal do Cear\'a (UFC), Departamento de F\'isica,\\ Campus do Pici, Fortaleza - CE, C.P. 6030, 60455-760 - Brazil.}


\date{\today}

\begin{abstract}
In this paper, we investigate the thermodynamic properties of the noncommutative Dirac oscillator with a permanent electric dipole moment in the presence of an electromagnetic field in contact with a heat bath. Using the canonical ensemble, we determine the properties for both relativistic and nonrelativistic cases through the \textit{Euler-MacLaurin} formula in the high temperatures regime. In particular, the main properties are: the Helmholtz free energy, the entropy, the mean energy, and the heat capacity. Next, we analyze via 2D graphs the behavior of the properties as a function of temperature. As a result, we note that the Helmholtz free energy decreases with the temperature and $\omega_\theta$, and increases with $\omega$, $\Tilde{\omega}$, $\omega_\eta$, where $\omega$ is the frequency of the oscillator, $\Tilde{\omega}$ is a type of cyclotron frequency, and $\omega_\theta$ and $\omega_\eta$ are the noncommutative frequencies of position and momentum. With respect to entropy, we note an increase with the temperature and $\omega_\theta$, and a decrease with $\omega$, $\Tilde{\omega}$, $\omega_\eta$. Now, with respect to mean energy, we note that such property increases linearly with the temperature, and their values for the relativistic case are twice that of the nonrelativistic case. As a direct consequence of this, the value of the heat capacity for the relativistic case is also twice that of the nonrelativistic case, and both are constants, thus satisfying the \textit{Dulong-Petit} law. Lastly, we also note that the electric field does not influence the properties in any way.
\end{abstract}


\maketitle

\section{Introduction}

The Dirac oscillator (DO) is an exactly soluble model introduced in the context of relativistic quantum mechanics for spin-1/2 massive fermions (Dirac fermions) \cite{Moshinsky,Martinez}. Such a model was developed by Moshinsky and Szczepaniak in 1989 and has been also considered as an interaction term for modeling quark confinement in quantum chromodynamics (QCD) \cite{Moshinsky,Martinez}. To obtain the DO, we need to insert into the free Dirac equation (DE) a nonminimal coupling given by: ${\bf p}\to{\bf p}-im_0\omega\beta {\bf r}$, where ${\bf p}$ is the momentum operator, $i$ is the imaginary unit, $m_0$ is the rest mass of fermion with an angular frequency $\omega$, $\beta$ is one of the Dirac matrices, and ${\bf r}$ is the position operator \cite{Moshinsky,Martinez}. In particular, in the nonrelativistic limit, the DO becomes the quantum harmonic oscillator (QHO) with a strong spin-orbit coupling. Since it was introduced in the literature, the DO has already been verified experimentally \cite{Franco}, and has also been widely studied (applied) in different physical problems. For instance, we have the Aharonov-Bohm-Coulomb system \cite{Oliveira1}, noninertial effects in a cosmic string spacetime \cite{Oliveira2,Oliveira3}, thermodynamics \cite{Pacheco1,Pacheco2,Gu}, graphene \cite{Quimbay}, effects of spin \cite{Andrade1,Andrade2}, Landau levels \cite{Guvendi}, quantum optical \cite{Bermudez}, noncommutative phase space \cite{Cai}, Aharonov–Casher effect \cite{Candemir}, etc.

The study of the physical properties of materials, and in special, the thermodynamic (thermal) properties, is of great interest in condensed matter physics, solid-state physics, and materials science \cite{Gaskell,DeHoff,Tester,Dolling,Muhlschlegel,Coleman,Eckert,Foiles,Anthony,Wang,Grimvall}. Indeed, the efforts expended to get the knowledge of such properties are justified as much by practical needs as by fundamental science \cite{Balandin}. Some examples of such practical relevance can be found in Refs. \cite{Balandin,Mounet,Shahil,Pop,Alofi,Che,Ruoff,Philip}, where the thermodynamic properties of graphene, diamond, graphite, carbon nanotubes, nanostructured carbon materials, and nanofluids were investigated. Besides, the study of such properties is also of great relevance from a theoretical point of view (for low and high-energy quantum systems). For example, in relativistic quantum mechanics, such theoretical relevance can be found in Refs. \cite{Pacheco1,Pacheco2,Boumali,Oliveira4,Oliveira5,Montakhab,Hassanabadi,Santos}, where we have the own DO, set of neutral Dirac fermions with a magnetic dipole moment, Aharonov-Bohm quantum rings, two-dimensional relativistic gas, Klein-Gordon oscillator, and the graphene; while in nonrelativistic quantum mechanics such relevance can be found in Refs. \cite{Groote,Donfack,Nammas,Najafi,Khordad,Kria,Fotue,Olendski}, where we have quantum pseudodots and dots, quantum wires, nanofibers, magneto-polarons, and quantum wells.

The present paper has as its goal to investigate the thermodynamic properties of the noncommutative Dirac oscillator (NCDO) with a permanent electric dipole moment (EDM) in the presence of an external electromagnetic field in (2+1)-dimensions. In particular, we determine the properties for both relativistic and nonrelativistic cases in the high temperatures regime, where we use the canonical ensemble as the thermal background to accommodate a set of noninteracting $N$-particles ($N$-DOs) in contact with a thermal bath (heat reservoir) at a constant temperature $T$ (thermodynamic equilibrium temperature). From the physical point of view, the canonical ensemble is a closed system characterized by a constant number of particles ($N=cte$) in a constant volume ($V=cte$) in which only heat and work can be exchanged between the partial system and the thermal bath, where the total system (isolated from the rest of the universe) is constituted by: partial system ($N$-particles) + thermal bath \cite{Greiner}. So, in order to perform the calculations, we use the \textit{Euler-MacLaurin} sum formula to construct the canonical partition function of the system. Next, we explicitly determine the (macroscopic) thermodynamic quantities of our interest, namely, the Helmholtz free energy, the entropy, the mean (internal) energy, and the heat capacity.

This paper is organized as follows. In Section \ref{sec2}, we determine explicitly the thermodynamic properties of the NCDO with EDM in the presence of an external electromagnetic field. In Section \ref{sec3}, we present the results and discussions, where we analyze via 2D graphs the behavior of the thermodynamic properties as a function of temperature. In Section \ref{conclusion}, we finish our work with the conclusion.

\section{Thermodynamic properties of the system \label{sec2}}

In this section, we calculate the relativistic and nonrelativistic thermodynamic properties of a $N$-particles system in contact with a thermal bath at temperature $T$. According to statistical thermodynamics (mechanics), these properties are the Helmholtz free energy $F(T,V,N)$, the entropy $S(T,V,N)$, the mean energy $U(T,V,N)$, and the heat capacity $C_V(T,V,N)$, where all are, together with $V$ and $N$, extensive variables (already $T$ is an intensive variable), being $F(T,V,N)$ and $U(T,V,N)$ also state functions (thermodynamic potentials), respectively. So, to determine such properties, we must first calculate the partition function, which in turn is calculated from the energy spectrum. 

\subsection{The relativistic case\label{subsec1}}

Let's start our discussion with the following relativistic energy spectrum for the NCDO with EDM in the presence of an external electromagnetic field \cite{Oliveira5}
\begin{equation}\label{spectrum}
E^{\chi}_{n,m,s}=U_{pot}+\chi m_0c^2\sqrt{1+\frac{2\hbar N}{m_0c^2}\left(1+s\frac{(\omega-\sigma\tilde{\omega})}{\omega_\theta}\right)(\omega-\sigma\tilde{\omega}+s\omega_\eta)},
\end{equation}
where
\begin{equation}\label{potentialenergy}
U_{pot}\equiv-d_f E_0=-\sigma\vert d_f\vert E_0, \ \ (\sigma=\pm 1),
\end{equation}
and
\begin{equation}\label{N}
N=N_{eff}\equiv\left[2n+1-\kappa+\Big|m+s\frac{1-\kappa}{2}\Big|-s\left(m+s\frac{1-\kappa}{2}\right)\right]\geq 0,
\end{equation}
being $\chi=\pm 1$ a parameter that describes the positive or negative energy states (DO or anti-DO), $N$ is an effective quantum number, being $n=0,1,2,\ldots$ the radial quantum number and $m=0,\pm 1,\pm 2,\ldots$ the magnetic quantum number, $U_{pot}$ is the (electric) potential energy, with $d_f$ being the EDM, $E_0$ is the strength of the electric field ($E_0=cte>0$), $\sigma=\pm 1$ is a parameter that describes a positive or negative EDM, $\omega>0$ is the angular frequency of the DO, $\Tilde{\omega}=\frac{\vert d_f\vert\Phi}{m_0}>0$ is a type of cyclotron frequency (angular velocity), being $\Phi$ a magnetic field density (magnetic field per length), $\omega_\theta=\frac{2\hbar}{m_0\theta}>0$ and $\omega_\eta=\frac{\eta}{2\hbar m_0}>0$ are the NC frequencies of position and momentum, being $\theta$ and $\eta$ the position and momentum NC parameters, $s=\pm 1$ is the spin parameter (describes the spin ``up'' or ``down''), and $\kappa=\pm 1$ is a parameter that describes the two components of the Dirac spinor, respectively. In particular, this spectrum has a characteristic that allows us to calculate the thermodynamic properties of the system, which is the fact that it has a degeneracy, finite or infinite (finite or infinite degenerate states) depending on the values (signs) of $s$ and $m$. However, only a finite degeneracy, and a spectrum with positive energy ($\chi=+1$), are allowed for the calculation of such properties. For instance, considering the maximal spectrum, which is for $s=+1$ (spin ``up'') with $m<0$, $\sigma=-1$ (negative EDM), and $\kappa=+1$, we have the following finitely degenerate spectrum for the NCDO \cite{Oliveira5}
\begin{equation}\label{spectrum2}
E_{n,M}=U_{pot}+m_0c^2\sqrt{1+\frac{4\hbar}{m_0c^2}\left(1+\frac{(\omega+\tilde{\omega})}{\omega_\theta}\right)(\omega+\tilde{\omega}+\omega_\eta)[n+M]}>0,
\end{equation}
where for simplicity we rename $m$ by $-M$, with $M\ge 1$.

Now, let's focus our attention on the fundamental object of the statistical mechanics for the canonical ensemble, the so-called partition function $Z$, which is defined as the sum of all possible macroscopic quantum states of the system \cite{Greiner}. Explicitly, the one-particle partition function ($N=1$) is given by the following expression \cite{Pacheco2,Greiner,Oliveira4}
\begin{equation}\label{partition}
Z(T,V,1)=\sum_{k=1}^\infty\Omega(E_k)e^{-\beta E_k},
\end{equation}
where $\beta=\frac{1}{k_B T}$ is the Boltzmann factor, with $k_B$ being the Boltzmann constant, and the quantity $\Omega(E_k)$ is the degree of degeneracy (or simply the degeneracy) for the energy level $E_k$ (number of microstates of the system with energy $E_k$), where the spectrum $E_k$ is given by
\begin{equation}\label{spectrum3}
E^{}_{k}=U_{pot}+m_0c^2\sqrt{1+ABk},
\end{equation}
where we define $k\equiv n+M\ge 1$, being $k$ a new quantum number, and $A\equiv\frac{4\hbar}{m_0c^2}=4\lambdabar$ ($\lambdabar$ is the reduced Compton wavelength) and $B\equiv\left(1+\frac{(\omega+\tilde{\omega})}{\omega_\theta}\right)(\omega+\tilde{\omega}+\omega_\eta)$. Thus, to determine $\Omega(E_k)$, it is necessary to take into account that for each quantum level (state) described for a specific pair ($n,M$), there are $2M+1$ different degenerate states \cite{Pacheco2,Oliveira4}. In this way, the total degree of degeneracy (total number of microstates) is given by the following equation
\begin{equation}\label{degeneracy}
\Omega(E_k)=\sum_{M=1}^k(2M+1)=k(k+2).
\end{equation}

Therefore, the partition function \eqref{partition} becomes
\begin{equation}\label{partition2}
Z(T,V,1)=\sum_{k=1}^\infty k(k+2)e^{-[\bar{\beta}+\tilde{\beta}\sqrt{1+ABk}]},
\end{equation}
where we define $\bar{\beta}\equiv\beta U_{pot}=\frac{\vert d_f\vert\Phi}{k_B T}\ge 0$ and $\tilde{\beta}\equiv\beta m_0 c^2=\frac{m_0 c^2}{k_B T}\ge 0$ (the sign $\ge$ will depend on how $T$ behaves). 

Before proceeding, it is advisable to analyze the convergence of the partition function \cite{Pacheco1,Pacheco2,Oliveira4}. So, the function $f(x)=x(x+2)e^{-[\bar{\beta}+\Tilde{\beta}\sqrt{1+ABx}]}$ is a monotonically decreasing function if the following associated integral
\begin{eqnarray}\label{integral}
I(\bar{\beta},\Tilde{\beta})&=&\int_{1}^{\infty}f(x)dx=\left[\frac{240}{(AB)^3\Tilde{\beta}^6}+\frac{240\sqrt{1+AB}}{(AB)^3\Tilde{\beta}^5}+\frac{(96+144AB)}{(AB)^3\Tilde{\beta}^4}\right]e^{-[\bar{\beta}+\Tilde{\beta}\sqrt{1+AB}]} \nonumber \\
&& +\left[\frac{(16+64AB)\sqrt{1+AB}}{(AB)^3\Tilde{\beta}^3}+\frac{(16+22AB)}{(AB)^2\Tilde{\beta}^2}+\frac{6\sqrt{1+AB}}{AB\Tilde{\beta}}\right]e^{-[\bar{\beta}+\Tilde{\beta}\sqrt{1+AB}]},
\end{eqnarray}
is finite (convergent). Thus, from the theorems of convergent series, this implies that the partition function \eqref{partition2} is also convergent.

On the other hand, although the partition function \eqref{partition2} is a convergent function, it cannot be calculated exactly in a closed form \cite{Greiner}. However, for high temperatures ($T\to\infty$ or $\beta\ll 1$) we can obtain good approximations \cite{Greiner}. In that way, a systematic expansion of \eqref{partition2} for large $T$ is possible with the use of the \textit{Euler-MacLaurin} (sum) formula, in which the objective is to calculate the integrals numerically. In particular, the \textit{Euler-MacLaurin} formula is given by \cite{Greiner}
\begin{equation}\label{partition3}
Z(T,V,1)=\sum_{k=1}^\infty f(k)=\frac{1}{2}f(1)+\int_{1}^{\infty}f(x) dx-\sum_{p=1}^{\infty}\frac{1}{(2p)!}B_{2p}f^{(2p-1)}(1),
\end{equation}
or explicitly, as
\begin{equation}\label{partition4}
 Z(T,V,1)=\sum_{k=1}^{\infty}f(k)=\frac{1}{2}f(1)+\int_{1}^{\infty}f(x) dx-\frac{1}{12}f'(1)+\frac{1}{720}f'''(1)-\ldots+,
\end{equation}
where $B_{2p}$ are the Bernoulli numbers.

So, using the information above, the partition function \eqref{partition4} will become
\begin{eqnarray}\label{partition5}
Z(T,V,1)&=&\left[\frac{3}{2}+\frac{240}{(AB)^3\Tilde{\beta}^6}+\frac{240\sqrt{1+AB}}{(AB)^3\Tilde{\beta}^5}+\frac{(96+144AB)}{(AB)^3\Tilde{\beta}^4}\right]e^{-[\bar{\beta}+\Tilde{\beta}\sqrt{1+AB}]} \nonumber \\
&& +\left[\frac{(16+64AB)\sqrt{1+AB}}{(AB)^3\Tilde{\beta}^3}+\frac{(16+22AB)}{(AB)^2\Tilde{\beta}^2}+\frac{6\sqrt{1+AB}}{AB\Tilde{\beta}}\right]e^{-[\bar{\beta}+\Tilde{\beta}\sqrt{1+AB}]} \nonumber \\
&& -\left[\frac{8\sqrt{1+AB}-3AB\Tilde{\beta}}{24\sqrt{1+AB}}+\frac{(AB)^3\Tilde{\beta}}{2880(1+AB)^{5/2}}-\frac{(AB)^2\Tilde{\beta}}{720(1+AB)^{3/2}}\right]e^{-[\bar{\beta}+\Tilde{\beta}\sqrt{1+AB}]} \nonumber \\
&&-\frac{1}{720}\left[\frac{AB\Tilde{\beta}}{\sqrt{1+AB}}+\frac{3(AB)^3\Tilde{\beta}^2}{8(1+AB)^{2}}-\frac{(AB)^2\Tilde{\beta}^2}{(1+AB)}+\frac{(AB)^3\Tilde{\beta}^3}{8(1+AB)^{3/2}}\right]e^{-[\bar{\beta}+\Tilde{\beta}\sqrt{1+AB}]}+O(\Tilde{\beta}^4). \nonumber \\
&&
\end{eqnarray}

Therefore, in the high temperatures regime, the partition function \eqref{partition5} takes the form
\begin{equation}\label{partition6}
Z(T,V,1)\simeq\left(\frac{240}{(AB)^3\Tilde{\beta}^6}\right),
\end{equation}
or for a set of $N$-particles, as
\begin{equation}\label{partition7}
Z(T,V,N)\simeq\left(\frac{240}{(AB)^3(m_0 c^2)^6\beta^6}\right)^N,
\end{equation}
where we use the fact that $\Tilde{\beta}=\beta m_0 c^2$. Before calculating the thermodynamic properties of interest, it is convenient to make a quick observation about the partition function \eqref{partition6}. For instance, in the absence of the EDM ($\Tilde{\omega}\to 0$), and of the NC phase space ($\omega_\eta\to 0$ and $\omega_\theta\to\infty$), we get exactly the partition function of the usual DO in (3+1)-dimensions \cite{Pacheco2}, i.e., \eqref{partition6} is a generalization of the particular case already obtained in the literature.

Now, let's concentrate on the main thermodynamic properties, which are the Helmholtz free energy, the entropy, the mean energy, and the heat capacity. Mathematically, these properties are written in the following form \cite{Pacheco2,Oliveira4,Greiner}
\begin{equation}\label{Helmholtz}
F(T,V,N)=-\frac{1}{\beta}\ \mathsf{ln}[Z(T,V,N)],
\end{equation}
\begin{equation}\label{entropy}
S(T,V,N)=k_B\beta^2\frac{\partial}{\partial\beta}F(T,V,N),
\end{equation}
\begin{equation}\label{energy}
U(T,V,N)=-\frac{\partial}{\partial\beta}\mathsf{ln}[Z(T,V,N)],
\end{equation}
\begin{equation}\label{capacity}
C_V(T,V,N)=-k_B\beta^2\frac{\partial}{\partial\beta}U(T,V,N).
\end{equation}

Then, using the partition function \eqref{partition7}, the properties above will be rewritten as
\begin{equation}\label{properties1}
\bar{F}\simeq -k_B T\ \mathsf{ln}\left(\frac{240k_B^6 T^6}{(AB)^3(m_0 c^2)^6}\right), \ \ \bar{S}\simeq\left[6+\mathsf{ln}\left(\frac{240k_B^6 T^6}{(AB)^3(m_0 c^2)^6}\right)\right], \ \ \bar{U}\simeq 6k_B T, \ \ \bar{C}_V\simeq 6k_B,
\end{equation}
where $\bar{F}=\frac{F}{N}$ (Helmholtz free energy per particle), $\bar{S}=\frac{S}{N}$ (entropy per particle), $\bar{U}=\frac{U}{N}$ (mean energy per particle), and $\bar{C}_V=\frac{C_V}{N}$ (heat capacity per particle), respectively. However, unlike $\bar{F}$, $\bar{U}$ and $\bar{S}$, here only $\bar{C}_V$. Besides, here the pressure is zero, i.e., $P=-\frac{\partial F}{\partial V}=0$.

\subsection{The nonrelativistic case\label{subsec2}}

Now, let's consider the thermodynamic properties for the nonrelativistic case, which is the case where most of the phenomena of condensed matter physics or solid-state physics occur. However, let's first introduce the nonrelativistic energy spectrum for the NCDO with EDM in the presence of an external electromagnetic field. According to Ref. \cite{Oliveira5}, this spectrum is given by
\begin{equation}\label{spectrum3}
\varepsilon_{n,m,s}=U_{pot}+\hbar N\left(1+s\frac{(\omega-\sigma\tilde{\omega})}{\omega_\theta}\right)(\omega-\sigma\tilde{\omega}+s\omega_\eta),
\end{equation}
and considering the maximal spectrum, for instance, for $s = +1$ (spin ``up'') with $m<0$, $\sigma=-1$ (negative EDM), and $\kappa=+1$, we have a finitely degenerate spectrum as [rubens]
\begin{equation}\label{spectrum4}
\varepsilon_{n,M}=U_{pot}+2\hbar\left(1+\frac{(\omega+\tilde{\omega})}{\omega_\theta}\right)(\omega+\tilde{\omega}+\omega_\eta)[n+M]>0,
\end{equation}
where for simplicity we rename $m$ by $-M$, with $M\ge 1$.

So, the one-particle partition function is written by the following expression \cite{Pacheco2,Oliveira4,Greiner}
\begin{equation}\label{partition7}
Z(T,V,1)=\sum_{k=1}^\infty\Omega(\varepsilon_k)e^{-\beta\varepsilon_k}, \ \ (\beta=1/k_B T),
\end{equation}
where $\Omega(\varepsilon_k)$ is the degeneracy (degree of degeneracy), given by $\Omega(\varepsilon_k)=k(k+2)$, i.e., the same of the relativistic case, and the spectrum $\varepsilon_k$ is given by
\begin{equation}\label{energy4}
\varepsilon_k=U_{pot}+\bar{A}Bk, \ \ k=n+M=1,2,3,\ldots,
\end{equation}
where we define $\bar{A}\equiv 2 \hbar$, and $B$ is defined as $B\equiv\left(1+\frac{(\omega+\tilde{\omega})}{\omega_\theta}\right)(\omega+\tilde{\omega}+\omega_\eta)$. In that way, we can rewrite the partition function \eqref{partition7} as 
\begin{equation}\label{partition8}
Z(T,V,1)=\sum_{k=1}^\infty k(k+2)e^{-[\bar{\beta}+\bar{\bar{\beta}}k]},
\end{equation}
where $\bar{\beta}=\beta U_{pot}$ and $\bar{\bar{\beta}}=\beta\bar{A}B$.

However, unlike the relativistic case, here we do not need to use the Euler-Maclaurin formula. Therefore, the series in \eqref{partition8} converges in the following expression
\begin{equation}\label{partition9}
Z(T,V,1)=e^{-[\bar{\beta}-\bar{\bar{\beta}}]}\frac{(3e^{\bar{\bar{\beta}}}-1)}{(e^{\bar{\bar{\beta}}}-1)^3}.
\end{equation}

Thus, expanding the expression above, and considering the high-temperature regime where $\beta\ll 1$, the partition function takes the form (for $N$-particles)
\begin{equation}\label{partition10}
Z(T,V,N)\simeq\left(\frac{2}{(\Bar{A}B)^3\beta^3}\right)^N.
\end{equation}

Then, using \eqref{partition10}, the nonrelativistic thermodynamics properties will be written as
\begin{equation}\label{properties2}
\bar{F}\simeq -k_B T\ \mathsf{ln}\left(\frac{2k_B^3 T^3}{(\Bar{A}B)^3}\right), \ \ \bar{S}\simeq\left[3+\mathsf{ln}\left(\frac{2k_B^3 T^3}{(\Bar{A}B)^3}\right)\right], \ \ \bar{U}\simeq 3k_B T, \ \ \bar{C}_V\simeq 3k_B,
\end{equation}
where $\bar{F}=\frac{F}{N}$, $\bar{S}=\frac{S}{N}$, $\bar{U}=\frac{U}{N}$, and $\bar{C}_V=\frac{C_V}{N}$.

\section{Results and discussions\label{sec3}}

\begin{figure}[!h]
\centering
\includegraphics[width=16cm]{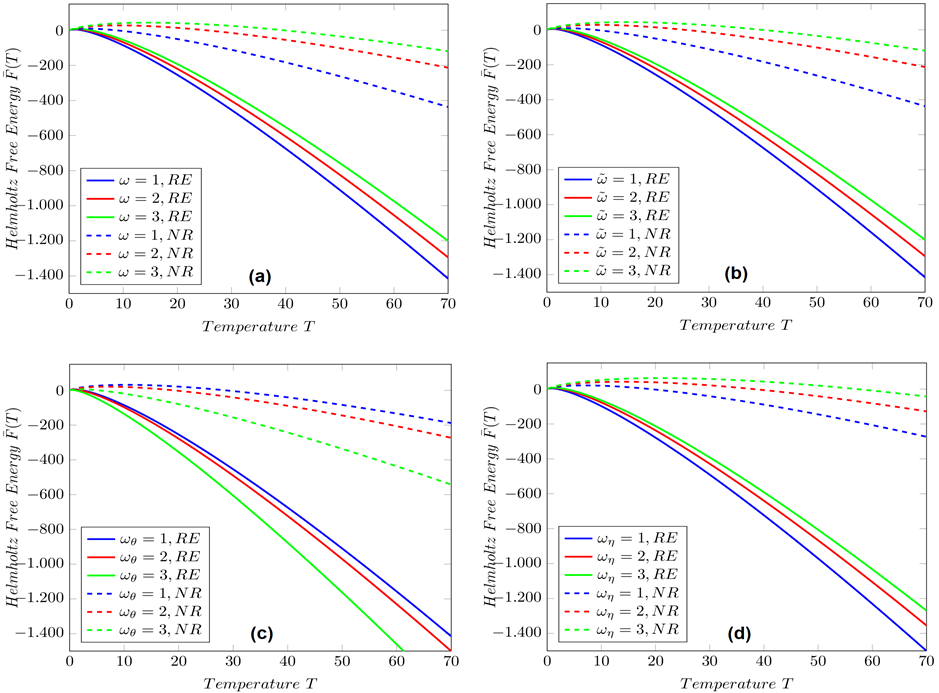}
\caption{Graph of $\bar{F}(T)$ versus $T$ for three different values of $\omega$, $\Tilde{\omega}$, $\omega_\theta$, and $\omega_\eta$, where the solid lines are for the relativistic (RE) case and the dashed lines are for the nonrelativistic (NR) case.}
\label{fig1}
\end{figure}

Now, we will discuss our (numerical) results via 2D graphs, where such graphs show the behavior of the thermodynamic properties as a function of temperature, both for the nonrelativistic and relativistic cases. For the sake of simplicity, here we also consider $\hbar=c=k_B=m_0=\vert d_f\vert=1$. So, in Fig. \ref{fig1} we have four graphs of Helmholtz free energy $\Bar{F}(T)$ as a function of temperature $T$ for three different values of $\omega$, $\Tilde{\omega}$, $\omega_\theta$, and $\omega_\eta$, which are the four frequencies of the system. For instance, in Fig. \ref{fig1}-(a) we have $\omega=\{1,2,3\}$ with $\Tilde{\omega}=\omega_\theta=\omega_\eta=1$, in Fig. \ref{fig1}-(b) we have $\Tilde{\omega}=\{1,2,3\}$ with $\omega=\omega_\theta=\omega_\eta=1$, in Fig. \ref{fig1}-(c) we have $\omega_\theta=\{1,2,3\}$ with $\omega=\tilde{\omega}=\omega_\eta=1$, and in Fig. \ref{fig1}-(d) we have $\omega_\eta=\{1,2,3\}$ with $\omega=\tilde{\omega}=\omega_\theta=1$ (these values are also used in the graphs of $\Bar{S}(T)$ versus $T$). 

In general, all graphs of Fig. \ref{fig1} behave in a similar way, that is, $\Bar{F}(T)$ decreases logarithmically (or monotonically) as a function of $T$. However, knowing that the variation or change of $\bar{F}$ can be written as $\Delta \bar{F}_{system}\le -\bar{W}$, where $\bar{W}=\bar{W}_{non-PV}\ge 0$ is a type of work done by the system (not mechanical work, or non-PV work, since $V=cte$ and $P=0$), implies that the smaller $\Delta \bar{F}_{system}$, the greater the value of $\bar{W}$ (the free energy of the system is depleted to do the work, i.e., is the energy available to do work). By way of comparison, $W$ is much larger for the relativistic case ($\bar{W}^{RE}\gg \bar{W}^{NR}$). On the other hand, since $\bar{F}$ can also be written as $\bar{F}\equiv\bar{U}-T\bar{S}$ ($\bar{U}\ge T\bar{S}$), it implies that a $\bar{F}$ with increasingly negative values, more entropy is generated ($\bar{S}$ increases), consequently, the system tends to equilibrium faster (faster heat exchange between the system and the thermal bath). Besides, with the exception of Fig. \ref{fig1}-(c), $\Bar{F}(T)$ increases with the increase of $\omega$, $\Tilde{\omega}$ and $\omega_\eta$ (for a fixed value of $T$). However, as $\Tilde{\omega}\propto \Phi$, $\omega_\eta\propto \eta$, and $\omega_\theta\propto 1/\theta$, implies that $\Bar{F}(T)$ increases with the increase of the magnetic field and of the NC parameters. It is important to mention that in the case of NC graphene, $\Bar{F}(T)$ also increases with the increase of the NC parameter $\eta$ \cite{Santos}. Also, the graphs of Figs. \ref{fig1}-(a) and \ref{fig1}-(b) are exactly the same. In fact, this is because $\omega$ and $\Tilde{\omega}$ are ``indistinguishable or interchangeable variables'' \cite{Oliveira5}.

\begin{figure}[!h]
\centering
\includegraphics[width=16cm]{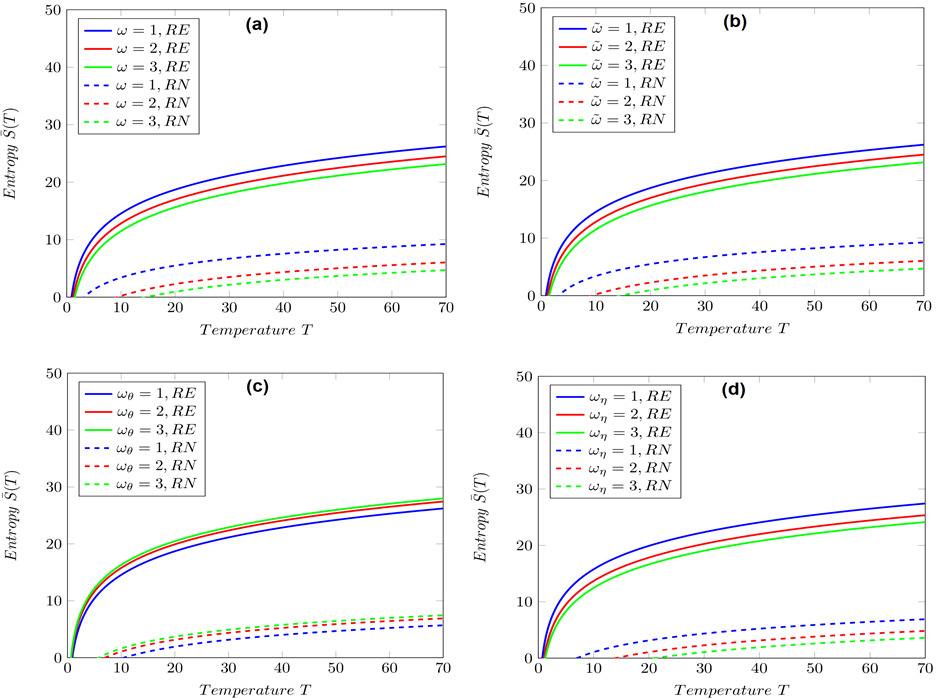}
\caption{Graph of $\bar{S}(T)$ versus $T$ for three different values of $\omega$, $\Tilde{\omega}$, $\omega_\theta$, and $\omega_\eta$, where the solid lines are for the relativistic (RE) case and the dashed lines are for the nonrelativistic (NR) case.}
\label{fig2}
\end{figure}

In Fig. \ref{fig2}, we have four graphs of entropy $\Bar{S}(T)$ as a function of temperature $T$ for three different values of $\omega$, $\Tilde{\omega}$, $\omega_\theta$, and $\omega_\eta$. In general, all these graphs behave in a similar way, that is, $\Bar{S}(T)$ increases logarithmically (monotonically) as a function of $T$ (as it should be, since $\Delta\Bar{S}=\Delta\Bar{S}_{system}\ge 0$). However, unlike of nonrelativistic case, $\Bar{S}(T)$ increases abruptly between $T=0$ and $T=10$. In fact, this occurs because the work $\bar{W}$ is much larger for the relativistic case ($\bar{W}^{RE}\gg\bar{W}^{NR}$), and therefore, the (change in) entropy is greater in the relativistic case. Still unlike the nonrelativistic case, in $T\simeq 0$ we have $\Delta\Bar{S}=0$. On the other hand, since $\Delta\bar{S}$ can also be written as $\Delta\bar{S}\ge\bar{Q}/T$, this implies that a large $\Delta\bar{S}$ requires a large $\bar{Q}$, i.e., the system must absorb more heat from the thermal bath (and for the relativistic case this absorption is greater). Besides, with the exception of Fig. \ref{fig2}-(c), $\Bar{S}(T)$ decreases with the increase of $\omega$, $\Tilde{\omega}$ and $\omega_\eta$ (for a given fixed value of $T$). However, as $\Tilde{\omega}\propto \Phi$, $\omega_\eta\propto \eta$, and $\omega_\theta\propto 1/\theta$, implies that $\Bar{S}(T)$ decreases with the increase of the magnetic field and of the NC parameters. It is important to mention that in the case of NC graphene, $\Bar{S}(T)$ also decreases with the increase of the NC parameter $\eta$ \cite{Santos}. In particular, this effect is expected because the noncommutativity leads to a decrease of the degeneracy in physical systems, reflecting in a reduction of the entropy in a NC phase space \cite{Santos}. So, analogous to Helmholtz free energy, the graphs of Figs. \ref{fig2}-(a) and \ref{fig2}-(b) are exactly the same.

\begin{figure}[!h]
\centering
\includegraphics[width=9cm]{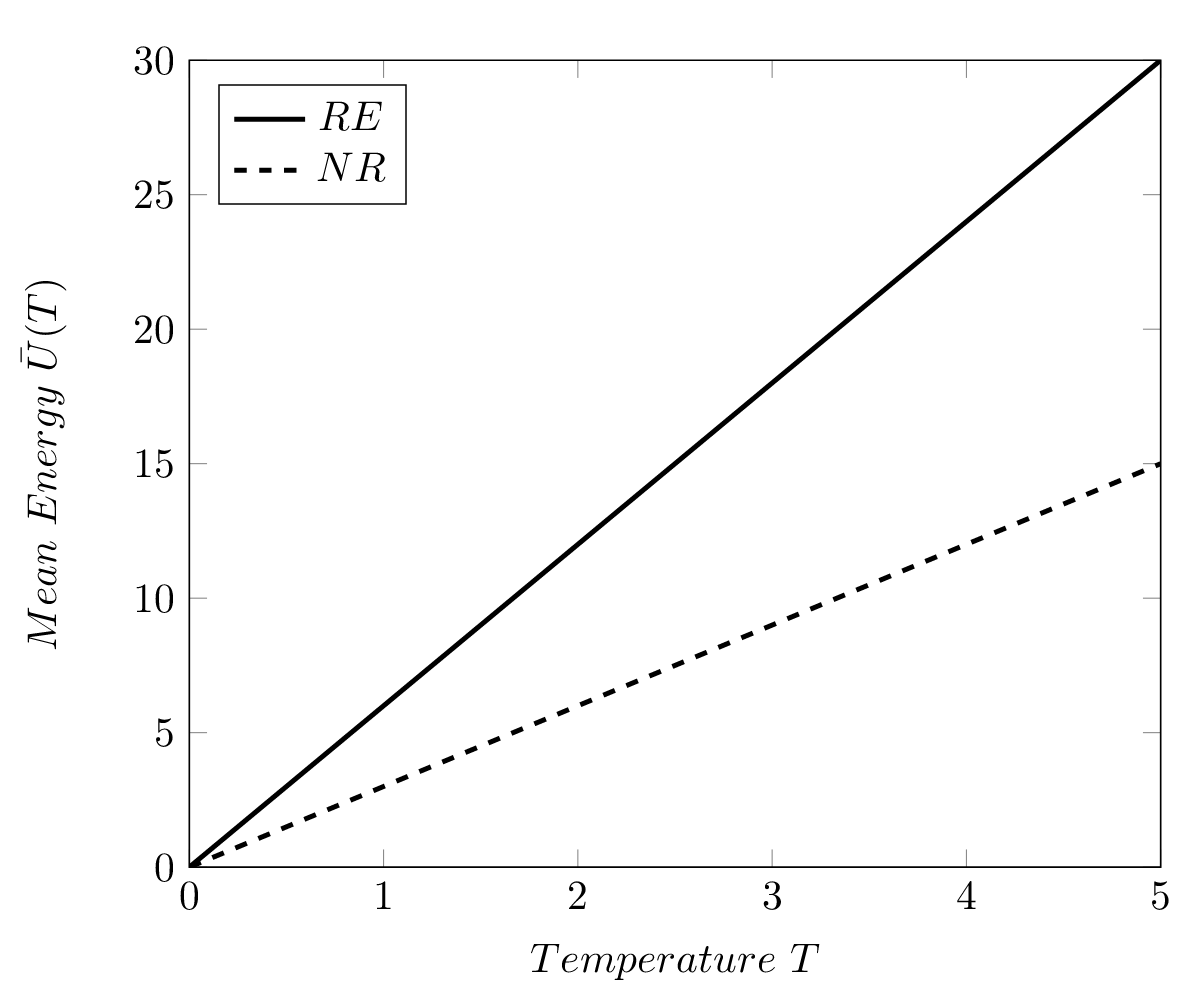}
\caption{Graph of $\bar{U}(T)$ versus $T$, where the solid line is for the relativistic (RE) case and the dashed line is for the nonrelativistic (NR) case.}
\label{fig3}
\end{figure}

Already in Fig. \ref{fig3}, we have the behaviour of the mean energy with the temperature, where shows $\Bar{U}$ as a linear function of temperature ($\Bar{U}$ increases linearly with $T$, i.e., only depends on temperature). In particular, we note that the value of $\Bar{U}$ for the relativistic case is twice of the nonrelativistic case, respectively. From a physical point of view, we can say that the total mean energy, given by $\Bar{U}=\Bar{U}_{pot}+\Bar{U}_{kin}$ (potential energy $+$ kinetic energy) of the $N$-particle or of the system (with random and disordered motion) is large for the relativistic case. Besides, as $\Bar{U}$ can also be written as $\Delta\Bar{U}=\bar{Q}-\bar{W}\ge 0$ (heat absorbed $-$ work done by the system), implies that the relativistic case absorbs much more heat and does much more work than the nonrelativistic case. Now, with regard to the heat capacity $\Bar{C}_V$, we note that such property is a constant that depends only on the Boltzmann constant $k_B$, and as a direct consequence of the mean energy, implies that the value of $\Bar{C}_V$ for the relativistic case is also twice of the nonrelativistic case. According to the literature \cite{Santos,Greiner,Oliveira4}, these results satisfy the \textit{Dulong-petit} law, which states that the heat capacity is a constant for high temperatures (or far from absolute zero). Moreover, the nonrelativistic case also agrees with such a law, because the heat capacity (of most crystalline solid substances) is $3k_B$ (per particle). Last but not least, it is worth mentioning that since the partition function in both nonrelativistic and relativistic cases does not depend on the electric field, implies that the thermodynamic properties also do not depend (as we have seen so far). In fact, there is no influence of the electric field on the thermodynamic properties because this field is not ``bound'' to the quantum number $k$ (as the four frequencies are).

\section{Conclusion\label{conclusion}}

In this paper, we investigate the thermodynamic properties of the NCDO with EDM in the presence of an external electromagnetic field. In this sense, we determine the properties for both relativistic and nonrelativistic cases in the high temperatures regime, where we use the canonical ensemble to a set of noninteracting $N$-particles ($N$-DOs) in contact with a thermal bath at a constant temperature. So, in order to perform the calculations, we use the \textit{Euler-MacLaurin} formula to construct the partition function of the system. Next, we determine explicitly the thermodynamic properties (per particle) of our interest, namely, the Helmholtz free energy $\bar{F}$, the entropy $\bar{S}$, the mean energy $\bar{U}$, and the heat capacity $\bar{C}_V$. So, comparing our relativistic partition function with another work, we verify that in the absence of the EDM and the NC phase space, we get the partition function of the literature.

Then, in order to better analyze our (numerical) results, we plotted in 2D graphs the behavior of thermodynamic properties as a function of temperature for three different values of $\omega$, $\Tilde{\omega}$, $\omega_\theta$, and $\omega_\eta$ (the four frequencies of the system). In that way, starting with the first thermodynamic property, we note that the graphs of $\Bar{F}$ vs. $T$, in general, have similar behavior, that is, $\Bar{F}$ decreases logarithmically (monotonically) as a function of $T$. Besides, $\Bar{F}$ increases with the increase of $\omega$, $\Tilde{\omega}$ (or magnetic field), and $\omega_\eta$ (or NC parameter $\eta$), and decrease with the increase of $\omega_\theta$. However, as $\omega_\theta\propto\frac{1}{\theta}$, it implies that $\Bar{F}$ increase with the increase of the NC parameter $\theta$. Besides, it is important to mention that the graphs for $\omega$ and $\Tilde{\omega}$ are exactly the same ($\omega$ and $\Tilde{\omega}$ are ``indistinguishable or interchangeable variables'').

Now, with respect to the second thermodynamic property, we note that the graphs of $\Bar{S}$ vs. $T$, in general, have similar behavior, that is, $\Bar{S}$ increases logarithmically (monotonically) as a function of $T$, and decrease with the increase of $\omega$, $\Tilde{\omega}$ (or magnetic field), and $\omega_\eta$ (or NC parameter $\eta$), and increase (decrease) with the increase of $\omega_\theta$ (NC parameter $\theta$). However, as $\omega_\theta\propto\frac{1}{\theta}$, it implies that $\Bar{S}$ decrease with the increase of the NC parameter $\theta$. On the other hand, unlike of nonrelativistic case, $\Bar{S}$ increases abruptly between $T=0$ and $T=10$. Besides, it is also important to mention that the graphs for $\omega$ and $\Tilde{\omega}$ are exactly the same ($\omega$ and $\Tilde{\omega}$ are ``indistinguishable or interchangeable variables'').

As for the third thermodynamic property, we note that the graphs of $\Bar{U}$ are linear functions of temperature ($\Bar{U}$ increases linearly with $T$). In particular, we note that the value of $\Bar{U}$ for the relativistic case is twice of the nonrelativistic case. Besides, unlike $\Bar{F}$ and $\Bar{S}$ (which depend on the four frequencies), $\Bar{U}$ only depends on temperature. Now, with respect to the last thermodynamic property, we have the heat capacity $\Bar{C}_V$, which is a constant that depends only on the Boltzmann constant $k_B$. As a direct consequence of the mean energy, implies that the value of $\Bar{C}_V$ for the relativistic case is also twice of the nonrelativistic case. According to the literature, these results satisfy the \textit{Dulong-petit} law, which states that the heat capacity is a constant for high temperatures. Last but not least, it is worth mentioning that since the partition function in both nonrelativistic and relativistic cases does not depend on the electric field, implies that the thermodynamic properties also do not depend.

\section*{Acknowledgments}

\hspace{0.5cm}The authors would like to thank the Conselho Nacional de Desenvolvimento Cient\'{\i}fico e Tecnol\'{o}gico (CNPq) for financial support.

\end{document}